\documentclass{esannV2}
\usepackage[dvips]{graphicx}
\usepackage[latin1]{inputenc}
\usepackage{amssymb,amsmath,array}

\usepackage{tabularx}
\newcolumntype{Y}{>{\centering\arraybackslash}X}

\usepackage{xcolor}

\usepackage{listings}
\usepackage{xcolor}
\lstdefinestyle{terminal}{
  basicstyle=\ttfamily\scriptsize,
  breaklines=true,
  frame=single,
  columns=fullflexible,
  keepspaces=true,
  showstringspaces=false,
}

\lstdefinestyle{pythoncode}{
  language=Python,
  basicstyle=\ttfamily\small,
  breaklines=true,
  frame=single,
  columns=fullflexible,
  keepspaces=true,
  showstringspaces=false,
}

\usepackage{appendix}

\usepackage{hyperref}

\begin{document}
\title{LLM-based Vulnerable Code Augmentation: Generate or Refactor?}

\author{Dyna Soumhane Ouchebara$^1$ and St\'ephane Dupont$^1$
%
\thanks{This study was funded by CyberExcellence project by Digital Wallonia, and benefited from computational resources made available on Lucia, the Tier-1 supercomputer of the Walloon Region, infrastructure funded by the Walloon Region under the grant agreement n\textdegree1910247.} 
%
\vspace{.3cm}\\
%
1- University of Mons - Computer science department \\
Mons - Belgium
}

\maketitle

\begin{abstract}
Vulnerability code-bases often suffer from severe imbalance, limiting the effectiveness of Deep Learning-based vulnerability classifiers. Data Augmentation could help solve this by mitigating the scarcity of under-represented vulnerability types. In this context, we investigate LLM-based augmentation for vulnerable functions, comparing controlled generation of new vulnerable samples with semantics-preserving refactoring of existing ones. Using Qwen2.5-Coder to produce augmented data and CodeBERT as a classifier on the SVEN dataset, we find that our approaches are indeed effective in enriching vulnerable code-bases through a simple process and with reasonable quality, and that a hybrid strategy best boosts vulnerability classifiers' performance. \href{https://github.com/DynaSoumhaneOuchebara/LLM-based-code-augmentation-Generate-or-Refactor-}{Code repository} is available.
\end{abstract}

\section{Introduction}

Deep learning models for software engineering tasks depend on large, diverse, and well-labeled datasets, yet collecting and annotating such data is often expensive and time-consuming. Data Augmentation (DA) mitigates data scarcity by synthetically expanding the training set, improving generalization and robustness without additional labeling effort. However, unlike in computer vision and NLP, DA for source code is challenging: code is a structured, executable artifact whose meaning relies on strict syntactic and semantic constraints, so small edits can easily break compilation or alter behavior. Effective Code Augmentation (CA) must therefore preserve syntax and semantics while allowing for enough diversity. This challenge is amplified in software vulnerability detection, where datasets are highly imbalanced both between vulnerable and safe classes and across vulnerability types (some being strongly under-represented). In this setting, DA can enrich minority classes with diverse yet semantically consistent variants, consequently reducing the impact of imbalance on vulnerability detection models. Recently, large language models (LLMs) have been widely adopted for code-related tasks such as generation, refactoring, documentation, and repair, offering a powerful new mechanism for code augmentation. Our work explores this line of research, and aims to answer the following questions:

\textit{\textbf{RQ1:} How effective is LLM-based Code Augmentation in enriching vulnerable code in highly imbalanced vulnerability code-bases?}

\textit{\textbf{RQ2:} Can LLM-based Code Augmentation boost the performance of Deep Learning models in the vulnerability classification task?}

\section{Related Work}

Zhuo et al. \cite{zhuo2023survey} categorize Code Augmentation methods into rule-based transformations (refactorings, renamings), model-based generation (GANs, pre-trained Transformers), and example interpolation methods (Mixup). Dong et al. \cite{dong2025nlp} directly apply NLP augmentation techniques to code and report improved performance of models, despite the risk of broken syntax. MIXCODE \cite{dong2023mixcode} generates semantics-preserving refactorings and then linearly mixes (Mixup) their embeddings to regularize code classifiers. BUGLAB \cite{allamanis2021buglab} uses a learned model that applies bug-inducing edits to benign code, producing synthetic buggy examples. In the vulnerability detection setting, MPDA \cite{mao2025mpda} augments imbalanced vulnerability datasets by combining classical oversampling methods, a GAN that synthesizes new vulnerable samples, and fuzzy sampling to augment minority-class instances. Qi et al. \cite{qi2024semantic} transform existing vulnerable and safe functions via handcrafted semantic-preserving edits to produce new variants that retain the original vulnerability labels. FGVulDet \cite{liu2024fgvuldet} employs an approach that only perturbs code regions deemed unrelated to the vulnerable zone, so that new samples share the same vulnerability while differing in surrounding context. VULGEN and VGX \cite{nong2023vulgen, nong2024vgx} mine vulnerability-introducing edit patterns from real patches and reapplies these patterns at predicted suitable locations in benign code to generate realistic vulnerabilities. More recently, Deng et al. \cite{deng2024lert} explored LLMs for vulnerable data generation. They prompt GPT4 to synthesize vulnerable versions of safe functions for under-represented CWE types, and to further self-filter the outputs based on quality checks. Our approach follows this LLM-based line but explores controlled generation and refactoring based on existing vulnerable functions from a reliable dataset to produce structurally and contextually diverse variants that preserve vulnerability semantics.

\section{Methodology}

\subsection{Generation-based data augmentation}
In the first generation strategy, we synthesize entirely new vulnerable functions using an instruction-tuned LLM. For each vulnerability type (CWE), we randomly select $m$ real-world functions from the training set and use them as examples in a few-shot prompt. The model is asked to generate $n$ new vulnerable functions per prompt, and this process is repeated $k$ times with different exemplar subsets, yielding $n \times k$ synthetic functions per vulnerability type.
\\~\\
Our \textbf{prompting scheme} relies on a fixed system message, which establishes the model's role and all global constraints, and a variable user message which provides instance-level information. The \textbf{system message} assigns the model the role of an \textit{expert in C/C++ and Python programming and software vulnerabilities} and enforces strict output constraints. The model must return C/C++ or Python code only, with no comments, explanations, markdown, or language tags, that must resemble real-world industrial C/C++ and Python and avoid toy identifiers. Function, variable, and type names are required to follow realistic project-style conventions. The \textbf{user message} begins with a description of the CWE. The model is instructed to generate $n$ independent function definitions, separated by markers such as ``func 1'', ``func 2'', etc. Structural constraints are imposed: each function must contain 20--150 non-empty lines, use project-like types, and include a vulnerability that is embedded in realistic logic rather than being the function's sole purpose. Finally, the prompt provides $m$ examples (from training set), and the model is instructed not to copy those but only to use them as guidance regarding naming conventions, code organization, and how the vulnerability typically appears in practice.
\\~\\
Once the functions are generated, we proceed to quality checks.
In the first place, we verify the \textbf{syntactical quality} of the generated samples by passing them through a C/C++ or Python parser. In a second place, we verify their \textbf{label quality}. Ideally, this would be conducted by security experts, but as a start for this research, we verified them using GPT-5.1 Thinking, which is one of the strongest LLMs available (at submission-time). For this, we give a randomly picked subset of $q$ generated functions to the LLM, and ask it to verify if it indeed contains a vulnerability of the given type or not.

\subsection{Refactoring-based data augmentation}
The second strategy produces augmented samples by refactoring functions already present in the dataset. For each vulnerability type and for every corresponding function, we prompt the LLM to generate $n$ refactored variants. We rely on \textbf{18 refactoring techniques} commonly used in prior work. These transformations alter code structure and surface form without changing semantics or the underlying vulnerability. This is a categorized list of techniques: Renaming (API, Arguments, Local variable and Method renaming), Adding unused elements (Arguments or Local variable adding), Dead code insertion (Dead for/if/if-else/switch/while adding or Duplication), Logic-preserving rewrites of control and expressions (For loop/If enhancement, Return optimal, Plus zero), Safety / robustness guards (Filed enhancement), Logging (Prints).
\\~\\
In terms of \textbf{prompting scheme}, the \textbf{system message} instructs the model to act as \textit{``an expert in C/C++ and Python programming, code refactoring, and software vulnerabilities''} and to generate $n$ refactored versions of the given function. It must preserve the function's semantics, parameter list, return type, and vulnerability. To avoid inadvertent vulnerability repair, the prompt forbids removing dangerous operations. The system message restricts the refactoring space to the 18 techniques listed above; each generated function must apply at least two distinct transformations. Output must consist solely of C/C++ or Python code with realistic identifiers and without comments or explanations. As for the \textbf{user message}, it defines the requested number of refactorings, the required output format, the vulnerability description, and the function to be refactored. No  examples are provided, making this a zero-shot prompting setup.
\\~\\
Regarding quality verifications, we first check the \textbf{syntactical quality} just like for the previous approach. Then, we verify their \textbf{refactoring quality} by asking GPT-5.1 Thinking to check whether the quality of a randomly picked subset of $q$ refactored versions complies with the constraints and expected complexity. We note that the label quality of generated samples is not verified, since it is inherently preserved by design in this approach.

\section{Experimental Setup}
To evaluate the performance of our proposed approaches, we chose to apply them on the \textbf{SVEN} \textbf{Dataset} \cite{he2023sven}. This dataset was created in 2023 by manually inspecting security-related commits from three prior benchmarks (BigVul, CrossVul and VUDENC) and only keeping those with no quality issues, and the most critical CWE types. The statistics of SVEN's training set are in Table \ref{tab:cwe_counts} (after splitting their training data into 80\% train and 20\% validation).

\begin{table}[h!]
  \centering
  \small
  \begin{tabularx}{\textwidth}{|Y|Y|Y|Y|Y|Y|Y|Y|Y|}
    \hline
    {cwe-89} & 
    {cwe-125} & 
    {cwe-78} & 
    {cwe-476} &
    {cwe-416} & 
    {cwe-22} & 
    {cwe-787} & 
    {cwe-79} &
    {cwe-190} \\
    \hline
    141 & 
    107 & 
    69 & 
    60 & 
    45 & 
    42 & 
    41 & 
    39 & 
    32 \\
    \hline
  \end{tabularx}
  \caption{Number of functions per CWE in SVEN Training set.}
  \label{tab:cwe_counts}
\end{table}

Concerning the \textbf{Models}, we choose \textbf{Qwen2.5-Coder-32B}\footnote{https://huggingface.co/Qwen/Qwen2.5-Coder-32B-Instruct} for augmented data generation, for its high rank in code LLM benchmarks\footnote{https://huggingface.co/spaces/bigcode/bigcode-models-leaderboard}\footnote{https://huggingface.co/spaces/bigcode/bigcodebench-leaderboard} (at the time we conducted our experiments) especially on C/C++ and Python, and its suitable size; and \textbf{CodeBERT} as the vulnerability classifier for its well-established code representation capabilities as well as its lightness (allowing for quick fine-tuning).
\\~\\
In terms of \textbf{Evaluation Metrics}, our augmentation approaches are assessed through the number of new vulnerable samples we can generate, the percentage of augmentation obtained for each class, the average time per generated sample, and the quality of augmented data (syntax, refactors, labels). As for the vulnerability classifier, we evaluate it by reporting the macro average f1 for an overall performance indication on all classes.
\\~\\
Our \textbf{Technical Setup} comprises 3 A100-SXM4 GPUs and a 16GB RAM.

\section{Evaluation}

\begin{table}[ht]
  \centering
  \footnotesize
  \begin{tabularx}{\textwidth}{|Y|Y|Y|Y|Y|Y|Y|}
    \hline
    Approach &
    N$^\circ$ of samples &
    \% of augmentation &
    Average time per sample &
    Syntax quality &
    Label quality &
    Refactor quality \\
    \hline
    Generation  & 
    3348 & 
    581\% & 
    13.38$s$& 
    100\% & 
    0\% & 
    /\\
    \hline
    Refactoring & 
    3348 & 
    581\% & 
    58.51$s$ & 
    84.5\% & 
    /& 
    100\%\\
    \hline
  \end{tabularx}
  \caption{Assessment metrics for our two augmentation approaches.}
  \label{tab:augmentation_metrics}
\end{table}

To answer \textit{\textbf{RQ1}}, we review Table \ref{tab:augmentation_metrics} which reports the assessment of our two augmentation approaches, which we used to generate 10 new functions at a time, stopping at a target of 500 functions in total per class. Using the Generation approach, we could increase the dataset size by 581\%, with an average speed of 13.38$s$ per generated sample, which looks reasonable under our resources. In terms of quality, all (100\%) samples proved syntactically correct. On the other hand, the label quality check, measured on a random subset of 10 generated functions per class, surprisingly gave a 0\% score. We, however, also checked the label quality of the original dataset and found 0\% for most CWEs. This means that: either the training data is too complex for GPT5.1 Thinking to find the vulnerabilities, or the dataset's label quality is questionable (which was not expected, since it has been manually curated by the authors). Further investigation will help us confirm our hypotheses. For the Refactoring approach, we could increase the dataset size by 581\% with a speed of 58.51$s$ per generated sample (much slower than the first approach). The syntax quality of 84.5\% correct functions is acceptable, and the refactoring quality (measured just like label quality above) is perfect.

\begin{table}[h!]
  \centering
  \small
  \begin{tabularx}{\textwidth}{|Y|Y|Y|Y|Y|}
    \hline
    Training data &
    Original data &
    Generation augmented &
    Refactoring augmented &
    Both augmentations \\
    \hline
    Macro F1 & 
    0.62 & 
    0.64 & 
    0.65 & 
    0.67 \\
    \hline
  \end{tabularx}
  \caption{Macro-average F1 score for different training data.}
  \label{tab:classification_metrics}
\end{table}

To answer \textit{\textbf{RQ2}}, we review Table \ref{tab:classification_metrics} which reports the performance of our vulnerability classifier before and after augmentation. The generation-based augmentation indeed improves the the classifier performance, with an overall Macro F1 score of 0.64 vs 0.62 on the original data. This increase is even more noticeable if we look at the minority classes, such as cwe-190 increased by 18\% and cwe-22 by 7\%. The refactoring-based augmentation gave a even slightly better improvement with 0.65 Macro F1, with a notable increase for cwe-022 by 18\% and cwe-416 by 7\%). Applying both augmentations proved to be the most helpful performance-wise, with an overall improvement of 5\% on Macro F1, and a clear boost for most CWEs (up to 24\% boost).

\section{Conclusion}
In this work, we proposed two LLM-based vulnerable code approaches: the first synthesizes entirely new vulnerable functions based on examples, and the second produces semantics and vulnerability preserving refactored versions of existing functions. Our experiments allowed us to answer our research questions. For \textit{\textbf{RQ1}}, we conclude that our LLM-based augmentation approaches are indeed effective in enriching vulnerable code-bases through a simple process and in a reasonable time, with great syntax and refactoring quality, though the label quality proved questionable. For \textit{\textbf{RQ2}}, we observe that LLM-based augmentation can indeed boost the performance of vulnerability classifiers, and deduce that the best strategy is a hybrid approach that applies both few-shot generation and refactoring.





\begin{footnotesize}

\end{footnotesize}


\newpage
\appendix 
\section{Detailed results}

Below are the detailed classification reports of our CodeBERT model using:

The original training data:

\begin{lstlisting}[style=terminal]
Classification Report:
              precision    recall  f1-score   support

     cwe-022       0.80      0.67      0.73         6
     cwe-078       0.91      0.91      0.91        11
     cwe-079       1.00      1.00      1.00         5
     cwe-089       1.00      1.00      1.00        20
     cwe-125       0.40      0.53      0.46        15
     cwe-190       0.33      0.20      0.25         5
     cwe-416       0.50      0.43      0.46         7
     cwe-476       0.43      0.38      0.40         8
     cwe-787       0.33      0.33      0.33         6

    accuracy                           0.67        83
   macro avg       0.63      0.61      0.62        83
weighted avg       0.68      0.67      0.67        83
\end{lstlisting}

The generation-based augmented data:

\begin{lstlisting}[style=terminal]
Classification Report:
              precision    recall  f1-score   support

     cwe-022       1.00      0.67      0.80         6
     cwe-078       1.00      1.00      1.00        11
     cwe-079       0.83      1.00      0.91         5
     cwe-089       1.00      1.00      1.00        20
     cwe-125       0.41      0.60      0.49        15
     cwe-190       0.50      0.40      0.44         5
     cwe-416       0.57      0.57      0.57         7
     cwe-476       0.40      0.25      0.31         8
     cwe-787       0.25      0.17      0.20         6

    accuracy                           0.70        83
   macro avg       0.66      0.63      0.64        83
weighted avg       0.70      0.70      0.69        83
\end{lstlisting}

The refactoring-based augmented data:

\begin{lstlisting}[style=terminal]
Classification Report:
              precision    recall  f1-score   support

     cwe-022       1.00      0.83      0.91         6
     cwe-078       1.00      1.00      1.00        11
     cwe-079       1.00      1.00      1.00         5
     cwe-089       1.00      1.00      1.00        20
     cwe-125       0.38      0.60      0.46        15
     cwe-190       0.00      0.00      0.00         5
     cwe-416       0.50      0.57      0.53         7
     cwe-476       0.60      0.38      0.46         8
     cwe-787       0.67      0.33      0.44         6

    accuracy                           0.71        83
   macro avg       0.68      0.63      0.65        83
weighted avg       0.72      0.71      0.70        83
\end{lstlisting}

\newpage
The data augmented with both approaches:

\begin{lstlisting}[style=terminal]
Classification Report:
              precision    recall  f1-score   support

     cwe-022       1.00      0.50      0.67         6
     cwe-078       0.92      1.00      0.96        11
     cwe-079       1.00      1.00      1.00         5
     cwe-089       1.00      1.00      1.00        20
     cwe-125       0.44      0.53      0.48        15
     cwe-190       0.40      0.40      0.40         5
     cwe-416       0.75      0.43      0.55         7
     cwe-476       0.38      0.38      0.38         8
     cwe-787       0.50      0.67      0.57         6

    accuracy                           0.71        83
   macro avg       0.71      0.66      0.67        83
weighted avg       0.73      0.71      0.71        83
\end{lstlisting}

\begin{figure}[h!]
\centering
\includegraphics[scale=0.43]{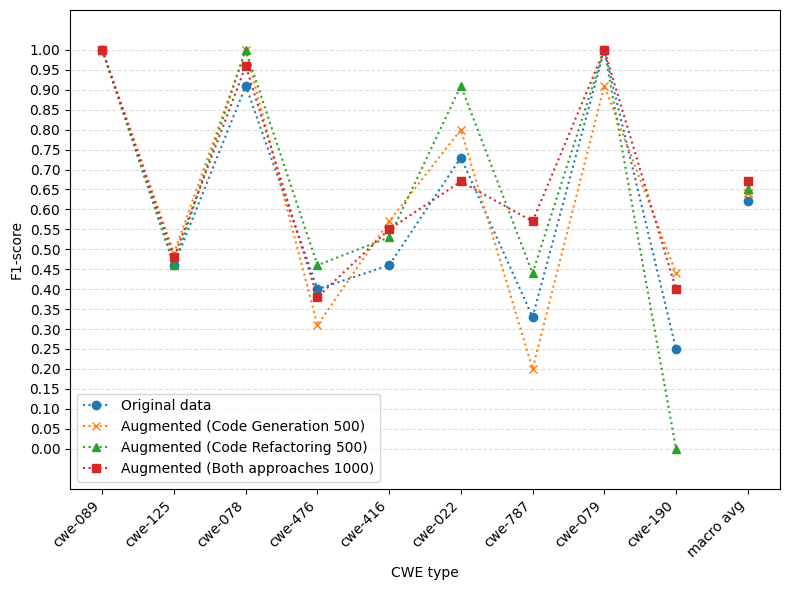}
\caption{F1-score per class for different training data.}\label{Fig:detailed_f1}
\end{figure}

To better interpret these results, Figure \ref{Fig:detailed_f1} gives a visual representation of the F1 score obtained per class, before and after each of the 3 augmentation approaches. The classes are ordered from the most represented to the least represented in the original training set (from left to right); this facilitates observation as to weather the classifier is less performant on least represented classes (those on the right) when trained on the original data, and if this performance gets better with augmentation.

We observe that the different augmentation approaches boost the classifier's capabilities for most classes, with a few exceptions (generation approach gives lower F1 for cwe-476 and cwe-787, refactoring approach gives lower F1 for cwe-190). Generation helps more than refactoring for 3 classes, while refactoring helps more for 5 classes. Applying both augmentations is not always better than applying only one of the two, but it ensures an overall (average on all classes) best classification performance. As for the performance on least represented classes, as expected, we do observe that the positive impact of augmentation is more noticeable on least representative classes, where we see a more important gap between the performance when training on the original dataset and when training on augmented datasets. However, we do not draw any clear pattern that would infer a better classification capability of most represented classes compared to the least represented ones (indeed the graphs do not have a right tailed shape).

This last observation lead us to investigate the existence of some other characteristics that would have an impact on the classification capability. The first one that came to mind is the length of the code snippets (in terms of n\textdegree of tokens in the snippet).

\begin{figure}[h!]
\centering
\includegraphics[scale=0.35]{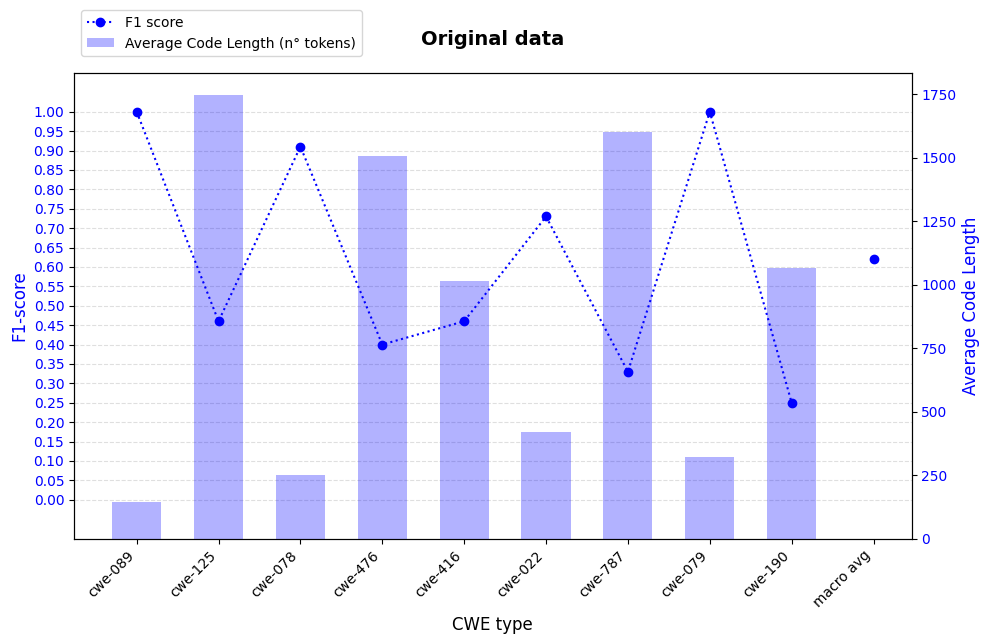}
\caption{F1 score vs Average code length on original data.}\label{Fig:f1_vs_codeLength}
\end{figure}

Figure \ref{Fig:f1_vs_codeLength} shows the relationship between the F1 score obtained for each vulnerability type and the average length (n\textdegree of tokens) of the code snippets from that type in the original unaugmented training data. As you can see, this investigation proved fruitful, as we can sense a pattern where CWEs with longer codes are more difficult to correctly classify than those with shorter codes (to each short (resp. tall) bar, representing a small (resp. big) code length, corresponds a high (resp. low) point in the graph, representing a high (resp. low) F1 score).

\begin{figure}[h!]
\centering
\includegraphics[scale=0.40]{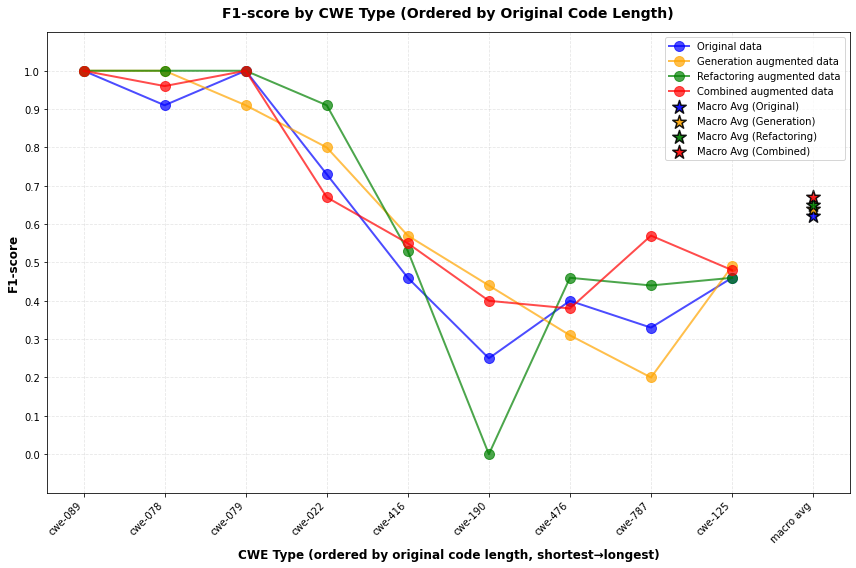}
\caption{F1 score vs Average code length on all datasets.}\label{Fig:f1_vs_codeLength_pattern}
\end{figure}

To better observe and confirm this pattern, Figure \ref{Fig:f1_vs_codeLength_pattern} shows the F1 score obtained per class over the 4 training datasets, with classes ordered this time according to the average length of their corresponding code snippets, from shorter on the left to longer on the right. Indeed, we can clearly see the pattern described earlier, where the classes with shorter codes (on the left) are much better classified than those with longer codes (on the right).

\begin{figure}[ht]
\centering
\includegraphics[scale=0.4]{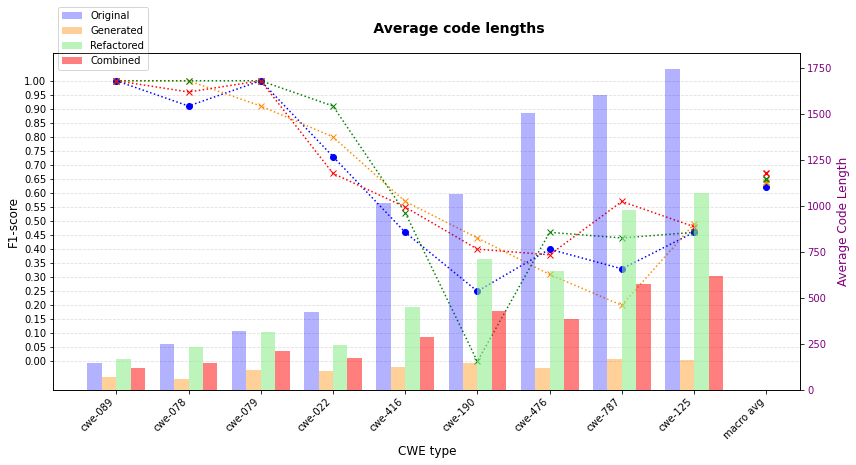}
\caption{Comparative average code lengths between the different training datasets.}\label{Fig:code_lengths}
\end{figure}

Since the length of code snippets seems to correlate with the classification performance, we now wonder about the length of the codes that were produced by our generation and refactoring prompts.
Figure \ref{Fig:code_lengths} reports the length between the original codes, those created using the generation approach and those using the refactoring approach (as well as the combination of both). We observe that the refactored codes are closer in length to the original codes, while the generated ones are sensibly smaller (despite having encouraged the model to generate longer codes through our prompt). It is important to note that length alone cannot measure similarity between codes, but it can at least give us some insight on it, so we can say that the refactored codes probably look more similar to the original codes, while the generated ones are far from resembling them. Further investigation is required to confirm this theory, and to make more informed conclusions about the comparative efficacy of both approaches.

\section{Prompts used}
Here we present the exact prompts we have used for the generation approach and the refactoring approach.

For the generation approach, here is the exact \textit{System Message} used :

\begin{lstlisting}[style=pythoncode]
"""
You are Qwen, created by Alibaba Cloud. You are a helpful assistant.
You are an expert in C and C++ and python programming and in security vulnerabilities.
General rules:
- Reply with C or C++ or python code only, no explanations, no comments, no markdown, no language tags.
- The code you generate must look like it belongs to a large real-world project, not like a textbook example.
- Function names, variable names and type names must look like they come from a mature C/C++ or python project and contains underscores , numbers and abbreviations.
- Avoid obvious, tutorial-style names.
"""
\end{lstlisting}

And below is the \textit{User Message} used :

\begin{lstlisting}[style=pythoncode]
"""
Vulnerability description:
{vulnerability_description}
You are given real-world vulnerable functions for this CWE. They come from large codebases 
(kernels, drivers, parsers, multimedia engines,etc). 
1) Output format:
Write {num_generatation_per_loop} independent function definitions (no main()) with seperators "func 1" "func 2" "func 3",etc.

NAMING AND STYLE REQUIREMENTS (VERY IMPORTANT):

3) Complexity:
- Each function should be at least 20 non-empty lines and at most 150.
- Use project-style types: structs, enums, typedefs (you can assume they exist),
    and realistic parameters (not just raw arrays).
- The vulnerability must be realistically embedded in the logic, not the only thing
the function does.

Here are example vulnerable functions of this vulnerability from real-world codebases.
Do NOT copy them. Use them as style and naming guidance:

{selected_examples}
"""
\end{lstlisting}
    
For the refactoring approach, here is the exact \textit{System Message} used :

\begin{lstlisting}[style=pythoncode]
"""
You are Qwen, created by Alibaba Cloud. You are a helpful assistant.
You are an expert in C and C++ and python programming, code refactoring, and security vulnerabilities.
 
Your job:
- Given ONE C/C++ or python function, generate several refactored versions of it.
- The refactored functions must:
  - Preserve the same high-level behavior and control-flow structure (same class label: VULNERABLE with a given vulnerability type).
  - Keep the same function signature: same parameter list and return type (function name may change unless explicitly told otherwise).
  - Keep all external side effects and return values equivalent for the same inputs.
  - Use different internal structure: local variable names, helper functions, loop forms, block structure, etc.
  - Preserve the same language as the input function.
 
Important constraints:
- Do NOT FIX OR REMOVE EXISTING VULNERABILITIES.
- If the input function is labeled as vulnerable, every refactored function must still contain the SAME underlying vulnerability type.
- If the input function is labeled as safe, all refactored versions must remain safe.
- You may move or reorder the code that causes the vulnerability, but you must NOT change its logical effect:
  - Do not change the arithmetic in buffer sizes and indices.
  - Do not add missing bounds checks or remove existing dangerous calls (e.g., memcpy/strcpy/strcat, unchecked pointer arithmetic, unsafe parsing, etc.).
  - Do not change which inputs lead to the vulnerable behavior.
 
Allowed refactoring techniques:
You may apply ONLY the following refactoring techniques (plus harmless superficial changes such as spacing or naming style). Any semantic change must still preserve behavior and vulnerability.
1) API renaming
   - Rename an API call to a synonym of its name.
2) Arguments adding
   - Add an unused argument to a function definition.
3) Arguments renaming
   - Rename an existing argument using a different, realistic identifier.
4) Dead for adding
   - Insert an unreachable for-loop at some location.
5) Dead if adding
   - Insert an unreachable if-statement at some location.
6) Dead if-else adding
   - Insert an unreachable if-else construct at some location.
7) Dead switch adding
   - Insert an unreachable switch statement at some location.
8) Dead while adding
   - Insert an unreachable while-loop at some location.
9) Duplication
   - Duplicate an existing assignment or simple statement and place the duplicate on the next line.
10) Filed enhancement
   - Add input-validation checks for None/null-like values around parameter use WITHOUT changing which calls lead to the vulnerability.
11) For loop enhancement
   - Rewrite for-loop bounds or layout in an equivalent way (e.g., different range expression) without changing the iteration set.
12) If enhancement
   - Rewrite an if-condition into an equivalent logical form (e.g., negation + inverted branches, reordered subconditions).
13) Local variable adding
   - Introduce an unused local variable.
14) Local variable renaming
   - Rename a local variable and update all its uses accordingly.
15) Method name renaming
   - Rename a function/method to a different realistic name while preserving its signature.
16) Plus zero
   - Modify a numeric expression by adding +0 (or an equivalent neutral term) so that the value is unchanged.
17) Print adding
   - Insert an extra print/logging call that does not remove or guard the vulnerability.
18) Return optimal
   - Rewrite the return expression as an equivalent variant (e.g., small conditional form) without changing returned values for any inputs.
 
Refactoring strength requirements:
- You must do more than simple renaming.
- DO NOT DUPLICATE THE SAME FUNCTION IN THE OUTPUT.
- For EACH refactored version, apply AT LEAST TWO distinct techniques from the list above, that can be applied on the given function.
- You may combine more than two techniques in one version.
- Do NOT explicitly mention which techniques you used.
- The output must be syntactically correct in the given language.
 
Code style constraints:
- C or C++ or python code only (no other languages).
- No comments.
- No explanations, no prose, no markdown, no language tags.
- Identifiers (function names, variables, structs) should look realistic and project-like (not toy examples).
- Do NOT simplify code into trivial textbook examples; keep roughly similar complexity and style as the original.
- Never mention vulnerabilities or security in the output.
- When asked for multiple refactors, output them in the exact format requested by the user.
"""
\end{lstlisting}

And below is the \textit{User Message} used:

\begin{lstlisting}[style=pythoncode]
"""
You will receive a single C/C++ or python function.

Your task is to generate {num_generatation_per_loop} independent refactored versions of that function,
following all rules from the system message.

Output format:
- Produce {num_generatation_per_loop} functions with seperators "func 1" "func 2" "func 3",etc.,
- Output code only, no comments, no explanations, no markdown.

Vulnerability description: {vulnerability_description}

The input function below is VULNERABLE with vulnerability type {cwe_name} and must remain vulnerable. Just rewrite and reorganize it
while keeping the same class label and preserving all existing bugs.

Here is the original function to refactor:

==== ORIGINAL FUNCTION START ====
{function}
==== ORIGINAL FUNCTION END ====
"""
\end{lstlisting}

\section{Models settings}
Table \ref{tab:parameters} reports the different parameters used to train/prompt our two models (QWEN for code augmentation, and CodeBERT for vulnerability classification).

\begin{table}[!h]
    \caption{Model parameters}
    \centering
    \begin{tabular}{|l|l|}
        \hline
        Parameter & Value \\
        \hline
        \textbf{CodeBERT classifier} & \\
        N\textdegree epochs & 15 \\
        Batch size & 16 \\
        Learning rate & $2 \times 10^{-5}$ \\
        Optimizer & Adam ($\beta_1$=0.9, $\beta_2$=0.999, $\epsilon=10^{-8}$) \\
        Max-length & 512 \\
        Random-state & 42 \\
        \textbf{Qwen2.5-Coder-32B generator} & \\
        Max-length & Default \\
        Max-new-tokens & Generation: 24000 \\
         & Refactoring: input-size $\times 10 + \epsilon ; \epsilon=400$ \\
        \hline
    \end{tabular}
    \label{tab:parameters}
\end{table}


\end{document}